\documentclass[aps,prx,reprint,superscriptaddress]{revtex4-2}

\usepackage{graphicx} 
\usepackage{longtable}
\usepackage{booktabs}
\usepackage[T1]{fontenc}
\usepackage{dcolumn}  
\usepackage[inline]{enumitem}

\usepackage{bm}        
\usepackage{amsfonts}  
\usepackage{amsmath}   
\usepackage{amssymb}  

\usepackage{amsthm}
\usepackage{tikz}
\usetikzlibrary{quantikz}
\usepackage{glossaries}

\usepackage{physics}

\usepackage{color}
\usepackage[colorlinks=true,allcolors=blue]{hyperref}

\newcommand{\pwisein}{\left\{ \begin{array}{ll}}
\newcommand{\pwiseout}{\end{array}\right.}

\newcommand{\avg}[1]{\langle#1\rangle}

\begin{document}

\title{Dissipative controller as a hidden source of errors in quantum gates}
\author{Serhat C. Kadıo{\u g}lu}
\affiliation{Department of Physics, Ko{\c c} University, Sarıyer, Istanbul 34450, Turkey}
\affiliation{School of Physics, Trinity College Dublin, College Green, Dublin 2, Ireland}
\author{Saulo V. Moreira}
\affiliation{School of Physics, Trinity College Dublin, College Green, Dublin 2, Ireland}
\author{Mark T. Mitchison}
\affiliation{School of Physics, Trinity College Dublin, College Green, Dublin 2, Ireland}
\affiliation{Department of Physics, King’s College London, Strand, London, WC2R 2LS, United Kingdom}
\author{Anthony Kiely}
\affiliation{School of Physics, University College Cork, College Road, Cork, Ireland}

\begin{abstract}
Quantum systems are typically controlled by classical devices that necessarily dissipate energy to the surrounding environment. By the fluctuation-dissipation theorem, this implies noise in the control field produced. We investigate the relationship between the resulting errors and the entropy produced by the macroscopic controller, working in the context of a concrete minimal model. In the most experimentally relevant regime of fast control and high fidelity, we find a tradeoff between dissipation and control errors. These findings are exemplified by perturbative calculations of the gate fidelity for simple gate implementations and of the fidelity of state transfer via shortcuts to adiabaticity in the Landau--Zener model. Our results quantitatively demonstrate that suppressing errors in quantum control typically comes at the cost of macroscopic dissipation, which is needed to tame fluctuations in the classical controller. 
\end{abstract}

    \maketitle 

\section{Introduction}

Experimental advances in realizing physical systems that can only be accurately described by the laws of quantum mechanics provide the background to the current interest in quantum technologies. These are new technologies which explicitly leverage the rules of quantum mechanics in their operation. The most famous are quantum computers, which can perform certain computations faster than a classical computer. The fundamental building block of the quantum computer is the quantum gate, analogous to a classical logical gate. As quantum devices are inherent very fragile, executing quantum gates requires high-fidelity control. Quantum control is a well studied field \cite{koch2022} with many distinctive methods and approaches including dynamical decoupling \cite{ezzell2023}, numerical optimization \cite{khaneja2005,Caneva2011,goerz2019}, composite pulses \cite{levitt1986}, and shortcuts to adiabaticity (STA) \cite{Gury-Odelin2019}.

In parallel, research on quantum thermodynamics \cite{vinjanampathy2016,goold2016,campbell2026} addresses questions of heat, work and entropy in the context of quantum systems. A key question for this field has been: What is the energetic cost of quantum control \cite{Horowitz2015, Auffves2022,Jaschke2023,Woods2023,Meier2025, Meier2026, Cepollaro2026} and can it be reduced \cite{abah2019,Carolan2022, FellousAsiani2023,Woods2024}?  This question has recently been investigated experimentally \cite{Cimini2020,Stevens2022}.
There has also been significant interest in potential energy--error trade-offs in quantum gates \cite{Campbell2017,stevens2025,fauquenot2025,Stevens2026,VanVu2024}. However, many previous works have addressed this problem at the level of the quantum system itself, whose energy scales are extremely small compared to the external control device such as a laser, electrode, or magnetic field generator. For this external device to behave as a classical controller, it must be a dissipative system undergoing rapid decoherence to suppress entanglement with its quantum target~\cite{Noh2008,Milburn2012,Layton2024}. Such dissipation not only implies an energy cost on the classical scale~\cite{Kiely2022}, but also necessarily entails fluctuations~\cite{Callen1951} that introduce errors into the evolution of the controlled quantum system. 

\begin{figure}[b]
    \centering
    \includegraphics[width=0.95\linewidth]{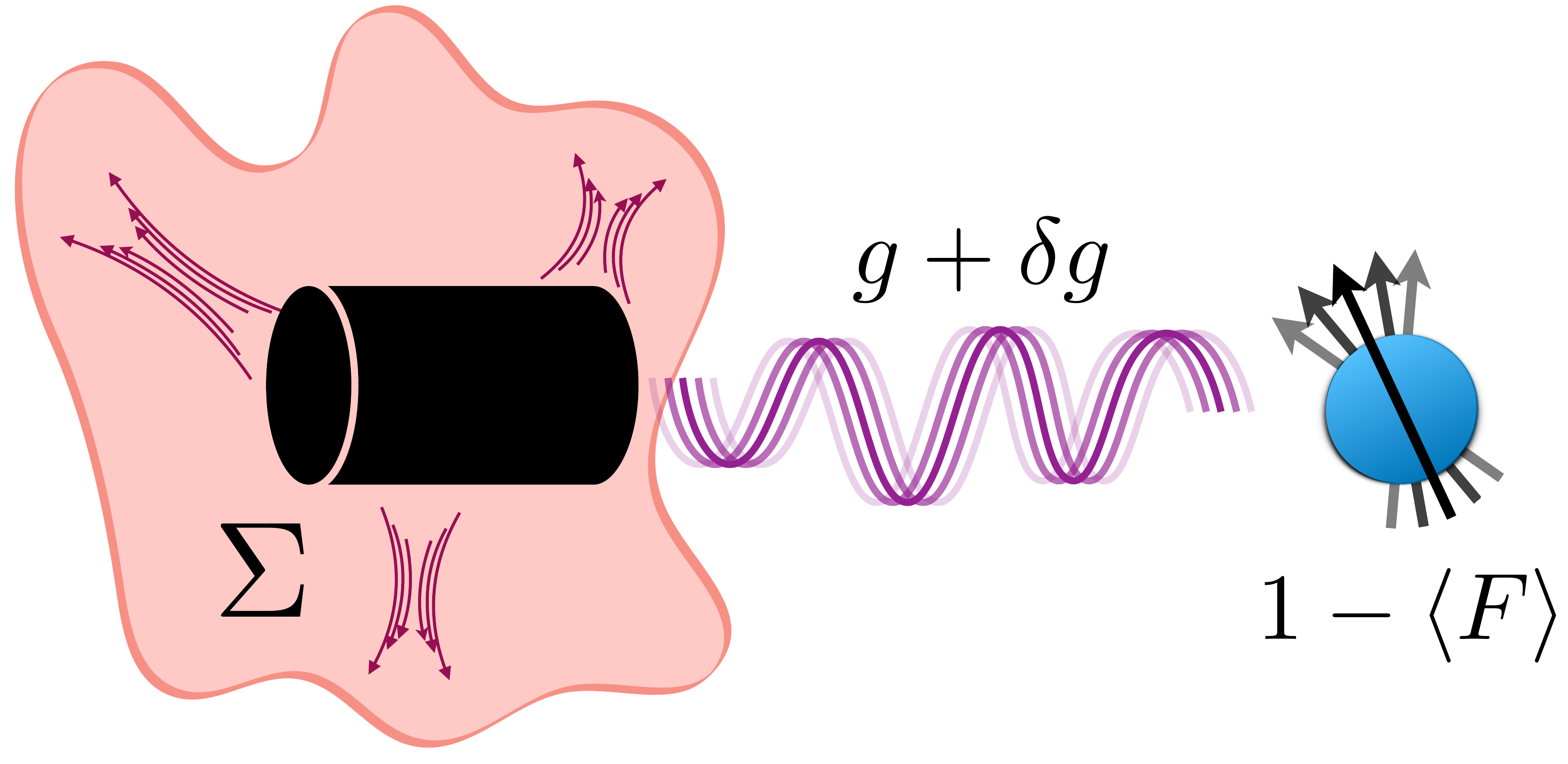}
    \caption{Schematic of setting considered. Classical control apparatus (black cylinder) dissipates into a surrounding thermal environment (red), quantified by the entropy production $\Sigma$. It is also used to generate the control field $g$ (purple lines), which has fluctuations $\delta g$. These fluctuations lead to average infidelity $\chi \approx 1-\avg{F}$ in the controlled quantum system (blue). The broad goal is to understand the interplay between $\Sigma$ and $\chi$.}
    \label{fig:diagram}
\end{figure}

In this work, we examine this interplay between energy cost and errors arising from the dissipative dynamics of the external control device. We describe the energetic cost at the macroscopic level using a classical model of the controller \cite{Torrontegui2017,tobalina2018,Kiely2022}, while characterizing the resulting errors microscopically through quantum dynamics, see Fig.~\ref{fig:diagram}. Specifically, we quantify the energy cost via the irreversible entropy production~\cite{landi2021} and the error from the fidelity loss due to the imperfect control signal implementation~\cite{Ruschhaupt2012,Goerz2014,Santos2017, Ball2016,Xuereb2023,Carolan2023,Poggi2024}. Unlike previous work on the dissipative cost of stabilizing non-equilibrium stationary states~\cite{Horowitz2015}, we consider time-dependent control protocols designed to prepare a target state or implement a target unitary, e.g., as relevant for quantum gate operations. Specifically, we focus on STA protocols, which have two advantages: (i) they allow for fast operations (in principle, faster than the characteristic error timescales) and (ii) they provide an analytical family of control solutions that achieve perfect fidelity in the absence of errors. This allows us to easily investigate how the particular control field profile affects both the dissipation and gate error.

In the following sections, we outline the macroscopic model of the dissipative control device (Sec.~\ref{sec:model}) and a perturbative description of the gate infidelity (Sec.~\ref{sec:fidelity}). We then discuss the simplest case of commuting controls, with an application to quantum gates, in Sec.~\ref{sec:commuting}. Finally, Sec.~\ref{sec:STA} considers the more complex scenario of quantum state transfer in the Landau--Zener model using STA. A discussion of our results and an outlook on future work is provided in Sec.~\ref{sec:conclusions}.

\section{Dissipative macroscopic controller}
\label{sec:model}
The control apparatus is modeled as a simple classical RL circuit,
following Ref.~\cite{Kiely2022}, in which the current flowing through a Helmholtz
coil generates the magnetic field $B(t)$ acting on the quantum system (e.g. a qubit). While we will focus on this specific model for concreteness, this approach is based on techniques for the stochastic thermodynamics of circuit
elements \cite{bruers2007,landi2013} and thus is extendable to more complex settings \cite{Freitas2020}.

The circuit
dynamics are described by the Langevin equation
\begin{equation}
2L\dot I =-RI+V(t)+\sqrt{2 R k_B T}\,\xi(t),
\end{equation}
where $2L$ is the total inductance of the coils, $R$ is the resistance, $k_B$ is Boltzmann's constant, $T$ is the temperature and $\xi(t)$ is Gaussian white noise representing the Johnson-Nyquist fluctuations.

For simplicity, we assume that the voltage required to implement a prescribed average control current $\langle I\rangle$ can be controlled perfectly as
\begin{eqnarray}
V(t) =2L \avg{\dot{I}}+ R \avg{I}.
\end{eqnarray} 
From this, the Langevin equation for the current fluctuations $\delta I = I-\avg{I}$ reduces to the equation for an Ornstein--Uhlenbeck process
\begin{eqnarray}
2L \delta\dot{I}= - R \delta I + \sqrt{2 R k_B T} \xi(t).
\end{eqnarray}
The stochastic fluctuations in the magnetic field generated are $\delta B = A \delta I$ where $A$ is a constant related to the coil geometry. These fluctuations in the control field, in turn, lead to decoherence~\cite{kiely2021}.

The corresponding entropy production, which can be physically related to Joule heating in the circuit,  is~\cite{Kiely2022}
\begin{equation}
\Sigma
=
\frac{k_B}{\gamma}
\int_0^\tau
dt\,
g(t)^2,
\label{entropyproduction}
\end{equation}
where $\tau$ is the total time of the control protocol and we have introduced the rate $\gamma=\mu^2A^2k_BT/(\hbar^2R)$ associated with the temperature, the dimensionless average control field $g(t)=\mu\avg{B(t)}/\hbar$, and the magnetic moment $\mu$ of the quantum system under control. In addition, we neglect a strictly positive transient term with an exponential decay time $\tau_c=2L/R$. While we have presented Eq.~\eqref{entropyproduction} for the entropy production in the context of the RL circuit, we stress that this result would also hold true for any setting where (i) the physical control field is odd under time reversal and (ii) the stochastic field fluctuations are well described by a Langevin equation $\delta \dot{g}=- \delta g/\tau_c+ \sqrt{2 \gamma}/\tau_c \xi(t)$.

In the next section, we investigate how current fluctuations, namely imperfections in the classical control apparatus, affect the quantum system. While the energetic cost introduced above belongs entirely to the classical control circuit and is described by classical Langevin dynamics, the resulting loss of accuracy is determined by the response of the quantum system. To quantify this effect, we treat the fluctuating control field as a classical stochastic perturbation acting on the quantum Hamiltonian. In other words, the energetic cost is attributed to the classical control apparatus, whereas its consequences are quantified through the microscopic dynamics of the quantum system.

\section{Infidelity of noisy quantum control}
\label{sec:fidelity}
We assume that the control protocols are designed to achieve perfect unit fidelity in the absence of the aforementioned fluctuations due to the dissipative controller. The presence of this noise therefore leads to an imperfect evolution in individual noise realizations and consequently reduces the average fidelity. We quantify this deviation from the ideal scenario using a second-order perturbative correction to the noise-averaged fidelity.

\subsection{Fidelity susceptibility}

We start by writing the system Hamiltonian as $H(t)=H_0(t)+ \hbar \, \delta g(t)W$ where $H_0(t)=H_S+ \hbar g(t) W$ denotes the noise-free Hamiltonian and $H_S$ is the static part of the Hamiltonian. $W$ is both the control operator associated with $g(t)$ and the noise operator for the stochastic fluctuations $\delta g(t)$ which are assumed small (see next subsection). We consider two different notions of fidelity: the state fidelity $F_S=|\langle \psi(\tau)|\psi_0(\tau)\rangle|^2$ between the final state $|\psi(\tau)\rangle$ and the target state $|\psi_0(\tau)\rangle$, and the unitary gate fidelity $F_U=|\operatorname{Tr}[U_0^\dagger(\tau)U(\tau)]|^2/d^2$ between the true evolution operator $U(\tau)$ and a target gate $U_0(\tau)$ ($d$ is the system's Hilbert-space dimension). The former describes the accuracy of specific state-preparation or state-transfer protocols, while the latter characterizes the fidelity of evolution in a state-independent way. In both cases, we assume the noise-free Hamiltonian $H_0(t)$ implements the target state $|\psi_0(\tau)\rangle$ or unitary $U_0(\tau)$ perfectly \cite{Poggi2024}.

The fidelity susceptibility $\chi \approx 1 - \avg{F}$, where $\langle F \rangle$ denotes the noise-averaged fidelity, is well known (see for example Ref.~\cite{Poggi2024}). While typically static noise is considered, here we
incorporate the temporal noise correlation function to describe time-dependent fluctuations \cite{green2013}. This allows us to write the susceptibility as
\begin{eqnarray}
\label{perturbative_susceptibility}
    \chi = 
\int_0^\tau dt\int_0^\tau ds\,
C(t,s)
K(t,s),
\end{eqnarray}
where the kernel $K(t,s)$ depends on the quantum system considered and we introduce the correlation
function of the control field fluctuations \begin{eqnarray}
\label{noise_correlation_function}
C(t,s)=\langle\delta g(t)\delta g(s)\rangle
&=&
\frac{\gamma}{\tau_c}
\exp\left[
-|t-s|/\tau_c
\right].
\end{eqnarray}
Note that the white noise limit is $\tau_c \rightarrow 0$, where $C(t,s) \rightarrow 2\gamma\delta(t-s)$.

For the case of state fidelity, $K(t,s)=\avg{W_H(t)W_H(s)}_0
-\avg{W_H(t)}_0\avg{W_H(s)}_0$, is a connected correlation function. The time evolution operator for ideal Hamiltonian $H_0(t)$ is $U_0(t)$, which allows us to define $W_H(t)=U_0^\dagger(t)W(t)U_0(t)$ as the operator $W$ in the Heisenberg picture. The notation $\langle . \rangle_0$ denotes the expected value taken over the initial state. For the case of unitary gate fidelity, we instead have $K(t,s)=\Tr\!\left[
W_H(t)W_H(s)
\right]/d$. We will use $\chi_S$ and $\chi_U$ for the state and unitary susceptibilities respectively. Unlike $\chi_S$, $\chi_U$ is independent of the input state. While $\chi_S$ quantifies the
sensitivity of a specified state-transfer process, $\chi_U$ measures the average deviation of the noisy evolution $U(\tau)$ from the target control operation $U_0(\tau)$.

\subsection{Regime of validity}

Here, we determine the conditions under which the perturbative approximation~\eqref{perturbative_susceptibility} is well justified and its predictions are meaningful. To do so, it is useful to make explicit two assumptions underlying the derivation of $\chi$ by treating the noise as a perturbation.
Firstly, we assume that the classical controller is operating in the steady state (at times much longer than $\tau_c$) so that transients can be neglected. Secondly, we assume that the magnitude of fluctuations $\delta g$ is much smaller than the maximum field intensity, $ g_{\text{max}}$. In this way,
\begin{eqnarray}
    g_{\text{max}}&\gg& \sqrt{\avg{\delta g(t) \delta g(s)}} \\
    &\gg& \sqrt{\frac{\gamma}{ \tau_c}}.
\end{eqnarray}
Since the maximum field intensity should scale inversely with the duration of the protocol $\tau$, we are left with the final condition,
\begin{eqnarray}\label{perturb_regime}
    \frac{\tau^2 \gamma}{\tau_c} \ll 1. \label{eq:condition}
\end{eqnarray}
Therefore, in order for the perturbative approach to be valid, Eq.~\eqref{perturb_regime} must be fulfilled. As we can see, this is more easily achieved by a combination of short protocol duration, low temperatures and long noise correlation time.

\subsection{Inequalities for the state fidelity}

The susceptibility $\chi_S$ depends on the initial state as well as the protocol. To gain some intuition, however, we can derive some general bounds that hold for any initial state. To achieve this, we decompose the noise-free time evolution operator in terms of the particular solutions of the Schr{\"o}dinger equation, $U_0(t)=\sum_k \ketbra{\psi_k(t)}{\psi_k(0)}$. This allows us to rewrite the kernel function as, $K_S(t,s)=\sum_{k \neq 0}G_k(t) G^{*}_k(s)$ with $G_{k}(s)= \bra{\psi_0(s)} W \ket{\psi_k(s)}$ tracking the perturbation induced transitions between orthogonal evolved states ~\cite{Santos2017,Ruschhaupt2012}. 
For a general stationary noise,  using the Wiener-Khinchin theorem we can express the two point correlation function in terms of the power spectral density $S(\omega)$ as
\begin{eqnarray}
C(t,s)=\kappa(t-s)=\int_{-\infty}^{\infty} d\omega \, S(\omega) \, e^{i \omega (t-s)} .
\end{eqnarray}
From this,  the sensitivity can be rewritten as
\begin{eqnarray}
\label{chi_from_spectral_density}
\chi_S =  \sum_{k \neq 0} \int_{-\infty}^{\infty} d\omega \, S(\omega) \left| g_k(\omega) \right|^2,
\end{eqnarray}
where $g_k(\omega)=\int_0^{\tau} dt \, G_k(t) e^{i \omega t}$. For our particular case, the power spectral density has the form of a Lorentzian:
\begin{equation}
    \label{lorentzian}
    S(\omega)= \frac{\gamma}{\pi} \frac{1}{1+ \tau_c^2 \omega^2}.
\end{equation}
It is also convenient to define the total noise power $S_p=\int_{-\infty}^{\infty}d\omega\,S(\omega)
=\kappa(0)=\gamma/\tau_c$.

Using the Cauchy–Schwarz inequality, we note that $\left| g_k(\omega) \right|^2 \leq \tau \int_0^{\tau} dt  \, \left| G_k(t) \right|^2 $.  From this, the noise sensitivity can be upper bounded as
\begin{equation}
\chi_S \leq \tau S_p \int_0^\tau dt K_S(t,t) =  \frac{\tau}{2 \tau_c} \chi_W,
\end{equation}
where $\chi_W$ is the fidelity susceptibility corresponding to pure white noise. This highlights both (i) that the fidelity susceptibility decreases linearly for short times $\tau$ and (ii) the worst case scenario is associated directly with the case of white noise.

We now aim to derive a lower bound on $\chi_S$. Schematically suppressing the sum
over $k$, multiplying $\chi_S$ by $S_p$ produces a product of the form
$
S_p\chi_S
\sim
\left(\int d\omega\,S(\omega)\right)
\left(\int d\omega\,S(\omega)|g_k(\omega)|^2\right).
$
Since both integrals contain the same nonnegative weight $S(\omega)$,
we can apply the Cauchy--Schwarz inequality once more:
\begin{eqnarray}
 S_p \int_{-\infty}^{\infty} d\omega \, S(\omega) \left| g_k(\omega) \right|^2 &\geq& \left| \int_{-\infty}^{\infty} d\omega \, S(\omega) g_k(\omega) \right|^2 \nonumber \\ 
&=&\left|  \int_0^{\tau} dt \,\kappa(t) G_k(t)  \right|^2 ,
\end{eqnarray}
where the second line follows since the inner product is the same in the time or frequency domain. Summing over the
individual $k$ contributions, we obtain
\begin{eqnarray}
\chi_S &\geq& \frac{1}{S_p} \sum_{k \neq 0} \left|  \int_0^{\tau} dt \,  \kappa(t) G_k(t)\right|^2 \\
&=& \frac{\gamma}{ \tau_c} \int_0^{\tau} dt \int_0^{\tau} ds  \,e^{- (t+s) /\tau_c} \, \avg{W_H(t)W_H(s)}^c_0 . \nonumber \\
\end{eqnarray}
STA type solutions, which enable perfect fidelity in the noise-free setting, give control fields which can be written in terms of $t'=t/\tau$ i.e. $\tilde{g}(t')=g(\tau t')$. Using this new notation then gives our final result as
\begin{eqnarray}
    \chi_S \geq \frac{\gamma \tau^2}{ \tau_c}   (\Delta Z)^2_0 ,
\end{eqnarray}
where the exponentially weighted operator $Z= \int_0^1 dt' \, \tilde{W}_H(t') e^{-\tau t' /\tau_c}$, and $(\Delta Z)^2_0$ is the variance of $Z$ evaluated on the initial state. Note that alternative lower bounds have been derived in \cite{muller2022}.

Note that, in general, by the Cauchy Schwartz inequality,  the entropy production can be lower bounded by the pulse area $\theta=  \int_0^\tau dt \, g(t)$ as $ \Sigma/k_B \geq \frac{\theta^2}{\gamma\tau}$.

Combining these two inequalities and recognizing that the product $\chi \Sigma$ is independent of $\gamma$ gives
\begin{eqnarray}
   \chi_S \Sigma/ k_B &\geq& \frac{\theta^2 \tau}{\tau_c}  (\Delta Z)^2_0 .
   \label{eq:qsigmamul}
\end{eqnarray}
In the best-case scenario, the initial state is an eigenstate of the
operator $Z$, in which case its variance, and hence the lower bound on
$\chi_S$, vanishes. In the worst-case scenario, the maximum variance is given by one quarter of the squared spectral width, according to Popoviciu's inequality.

We now have all the tools needed to investigate the relationship between $\chi$ and $\Sigma$. It is important to note that different implementations of the same logical operation may be generated by different time-dependent Hamiltonians. Although they achieve the same ideal gate fidelity, they generally produce different Heisenberg evolutions of the noise operator, leading to different entropy production $\Sigma$ and fidelity susceptibilities $\chi$ in general. In the next section, we will start with the simplest setting.

\section{Special case of commuting dynamics}
\label{sec:commuting}

So far, we have derived both the state and gate error induced by controller fluctuations, and the corresponding entropy production. Now, we start to investigate the relation between these two quantities. We begin with the simplest case where $[H_S,W]=0$. This means that the noise free Hamiltonian commutes with the control operator $W$ for all times i.e. $[H_0(t),W]=0$. As a result, $W_H(t) =  W$ is a constant and the relevant kernels simplify to 
\begin{eqnarray}
    K_S(t,s) &=& (\Delta W)^2_0, \\
    K_U(t,s) &=& \Tr[W^2]/d.
\end{eqnarray}
The fidelity susceptibility then becomes
\begin{eqnarray}
    \chi &=&2\gamma K   \left[\tau_c
   \left(e^{-\frac{\tau}{\tau_c}}-1\right)+\tau \right],
   \label{eq:sus}
\end{eqnarray}
which is independent of the control field profile. For short protocol times $\tau$ it goes to zero, while for $\tau \gg \tau_c$, it scales as $\chi \propto \tau$.

Combining the exact susceptibility in Eq.~\eqref{eq:sus}
with the pulse-area lower bound on entropy production gives
\begin{eqnarray}
    \chi \Sigma/ k_B \geq K \theta^2 F(\tau/\tau_c) ,
\end{eqnarray}
where $F(x)=2 \left(\frac{e^{-x}-1}{x}+1\right)$ is an increasing function upper bounded by $2$.

In the following examples, we assume no static field, $H_S=0$, and write the driving field as $g(t)=\theta \dot{f}(t)$, with $f(0)=0$ and $f(\tau)=1$. The gate is then simply $U_0(\tau)  = \exp\left[ -i\theta W \right]$, and the energetic cost is
\begin{equation}
    \frac{\Sigma}{k_{\mathrm B}}
    =
    \frac{\theta^2}{\gamma}
    \int_0^\tau \mathrm{d}t\,[\dot{f}(t)]^2.
    \label{eq:gate_cost_angle}
\end{equation}
It is clear that $\Sigma \propto \frac{k_B \theta^2}{\gamma \tau}$. For example, the linear protocol $f(t)=t/\tau$ saturates the preceding lower bound, yielding $ \frac{\Sigma}{k_{\mathrm B}}=\frac{\theta^2}{\gamma\tau}$. 

In our numerical calculations, we instead employ the fifth-order polynomial protocol~\cite{Kiely2022}
\begin{equation}
    f(t)=10\left(\frac{t}{\tau}\right)^3-15\left(\frac{t}{\tau}\right)^4+6 \left(\frac{t}{\tau}\right)^5,
    \label{eq:gate_polynomial_protocol}
\end{equation}
which has vanishing first and second derivatives at both boundaries, thereby ensuring a smooth switching on and off of the control field. Polynomial ramps, including fifth-order forms, are commonly used in STA protocols because they conveniently satisfy the required boundary conditions \cite{Gury-Odelin2019,Kiely2022}, while periodic or sinusoidal pulse shapes are also common choices in quantum-gate and quantum-control implementations
\cite{abah2019,deffner2021,Bonana2018}. Here, we use the same fifth-order temporal profile for the gate and
Landau--Zener STA protocols, allowing the gate and state-transfer examples to be compared under closely matched control conditions. The resulting entropy production is
\begin{equation}
    \frac{\Sigma}{k_{\mathrm B}}
    =
    \frac{10\theta^2}{7\gamma\tau},
    \label{eq:gate_cost_polynomial}
\end{equation}
which is naturally larger than the linear protocol due to the increased control complexity.

\begin{figure}
\centering
\includegraphics[width=0.95\linewidth]{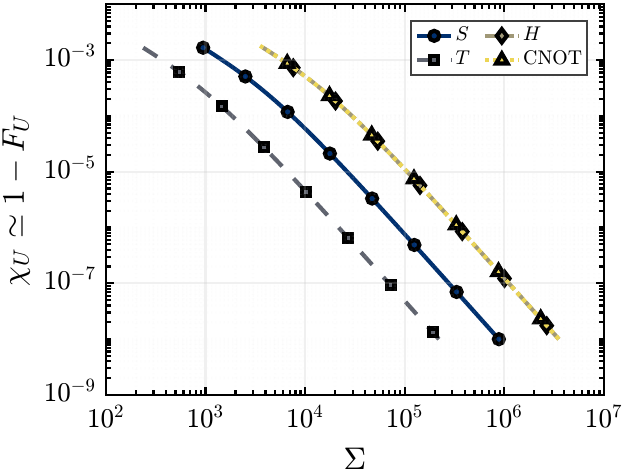}
    \caption{Gate infidelity $\chi_U\simeq1-F_U$ as a function of the
dissipative cost $\Sigma$ for the gates
$\{S,T,H,\mathrm{CNOT}\}$. 
The points along each curve are generated by varying $\tau/\tau_c$. Shorter durations correspond to larger dissipative costs and smaller gate susceptibilities. At a fixed duration, the gates share the same normalized susceptibility, while their energetic costs differ through the rotation angle $\theta$. For visual clarity,
markers are displayed only for a subset of the calculated points. The connecting lines represent the complete numerical dataset and are not fitted curves. }

\label{fig:gate_infidelity_energy}

\end{figure}

We evaluate the gate susceptibility $\chi_U$ for the single-qubit gates $U\in\{S,T,H\}$ and for the two-qubit CNOT gate over a range of protocol durations $\tau$, as shown in
Fig.~\ref{fig:gate_infidelity_energy}. Different gates incur different dissipative costs through their
rotation angles $\theta$, see Table \ref{tab:angles}. Thus, the resulting energy--error curves
separate horizontally, while sharing the same gate-infidelity dependence on
the protocol duration. We note that qualitatively similar results have been recently obtained by Stevens \& Deffner~\cite{Stevens2026}, albeit using a different noise model.

The CNOT gate provides a useful example for discussing how the
energetic cost should be treated for multi-qubit control operations.
We adopt the simple effective CNOT generator considered in
Refs.~\cite{Cavalcante2025,Santos2017}. Since all terms in this
generator share the same temporal profile, $g(t)$, the cost retains
the scaling $\int dt\,g^2(t)$ within the present effective model.
In a physical implementation, however, the same logical gate may
require several local drives and tunable interactions generated by
distinct control circuits. Consequently, the entropy production may
acquire a hardware-dependent prefactor that cannot be inferred from
the simplified effective generator alone. Experimental implementations
of two-qubit gates in superconducting circuits illustrate such
architecture-dependent control requirements
\cite{Rigetti2010,Chow2011,Sundaresan2020}. In particular,
Ref.~\cite{Dogan2023} presents a closely related effective Hamiltonian
whose terms are driven by a common time-dependent control function.
A quantitative determination of the corresponding prefactor would
require a microscopic model of the control hardware and is therefore
beyond the scope of this work.

\begin{table}
    \centering
    \caption{Control operators $W$ and total rotation angles $\theta$ used to
generate the gates considered in this work. Global phases
are omitted, and the operators are normalized so that
$K_U=1$.}
    \label{tab:gate_generators}
    \begin{ruledtabular}
\begin{tabular}{ccc}
Gate $G$ & Control operator $W$ & Rotation angle $\theta$ \\
\hline
$S$ &
$\sigma_z$ &
$\pi/4$ \\

$T$ &
$\sigma_z$ &
$\pi/8$ \\

$H$ &
$(\sigma_x+\sigma_z)/\sqrt{2}$ &
$\pi/2$ \\

CNOT &
$-\frac{1}{2}
(\sigma_1^z-\mathbb{I})
(\sigma_2^x-\mathbb{I})$ &
$\pi/2$
\end{tabular}
\end{ruledtabular}
\label{tab:angles}
\end{table}

\section{Landau--Zener protocol}
\label{sec:STA}
We now turn to the fidelity of state preparation using an STA, working in the context of the Landau--Zener (LZ) model. There are several motivations for considering this example. Landau--Zener-type dynamics and related control techniques have been employed in quantum-gate implementations, spanning nonadiabatic, adiabatic, and shortcut-to-adiabaticity protocols \cite{Li2019,Santos2015,Carolan2023,Cavalcante2025}. Moreover, the effective Hamiltonian for a two-qubit CNOT implementation reduces to a one-qubit LZ Hamiltonian within the relevant subspace \cite{Cavalcante2025}. Finally, since the LZ model admits an exact STA solution, the perturbative error formalism can be developed analytically despite the non-commuting nature of the dynamics.

The LZ Hamiltonian is given by
\begin{equation}
H_0(t)
=
\Delta \sigma_x
+
g(t)\sigma_z
+
G_{\rm CD}(t)\sigma_y ,
\end{equation}
where the counterdiabatic field
\begin{equation}
G_{\rm CD}(t)
=
-
\frac{\Delta \dot g(t)}
{2\left[\Delta^2+g(t)^2\right]},
\end{equation}
ensures perfect adiabatic evolution. There are two control fields, coupled to $\sigma_z$ and
$\sigma_y$, respectively. We assume that their fluctuations are
uncorrelated and treat their entropy-production and error
contributions separately. Denoting the $g(t)\sigma_z$ channel
by $\lambda$ and the $G_{\rm CD}(t)\sigma_y$ channel by $\beta$,
the corresponding perturbation operators are
$W_\lambda=\sigma_z$ and $W_\beta=\sigma_y$. Their Heisenberg-picture representations determine the respective two-time connected correlation kernels. Assuming identical noise correlation functions for the two independent control channels, the total kernel is given by

\begin{align}
K_S(t,s)
={}&K_S^{(\lambda)}(t,s)+K_S^{(\beta)}(t,s)
\label{eq:Cxy}
\end{align}
The explicit expressions for these kernels and their derivation are provided in Appendix \ref{app:STA}.

The system is initialized in the ground state of the bare Hamiltonian $H_{\mathrm{LZ}}(0)$, i.e., $H_0(0)$ without the counterdiabatic term. We consider the standard symmetric Landau--Zener sweep, starting from $g(0)=-g_{\max}$ and ending at $g(\tau)=+g_{\max}$. For the numerical calculations, we set $\Delta=0.5$ and $g_{\max}=5$.

\begin{figure}[t]
    \centering
    \includegraphics[width=0.95\linewidth]{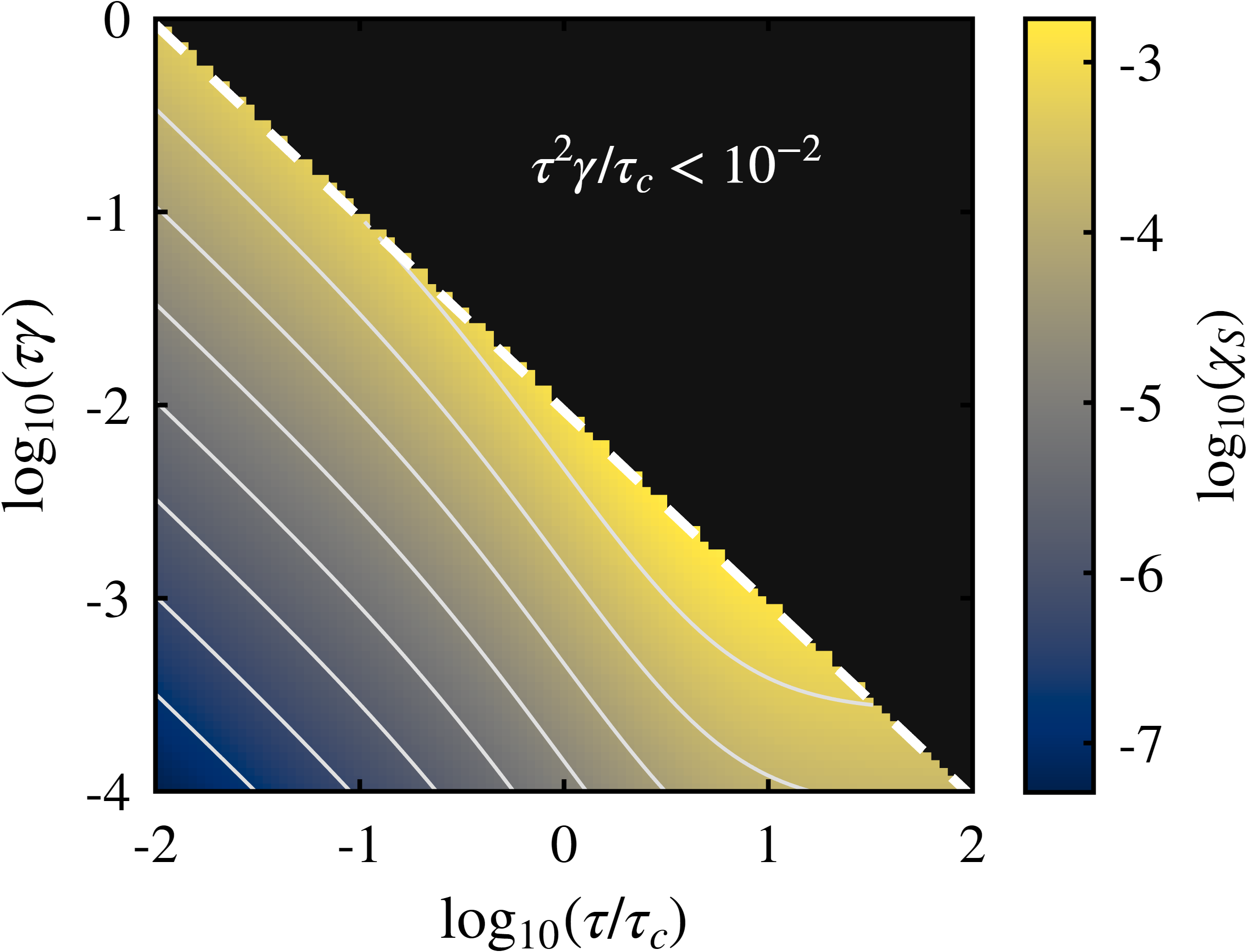}
    \caption{The fidelity susceptibility $\chi_S$ versus $\gamma$ and $1/\tau_c$. The results are calculated with a $5^{\rm th}$ order polynomial ramp control. The white dashed line represents the perturbative boundary $\tau^2\gamma/\tau_c<10^{-2}$.}
    \label{fig:q_heatmap}
\end{figure}

In this case, the total entropy production can be split into contributions from the
 control field and the counterdiabatic field as
\begin{eqnarray}
    \Sigma_{\mathrm{total}}
   & =&
    \Sigma_z
    +
    \Sigma_{\mathrm{CD}} \\
    &=&  
    \frac{k_B}{\gamma}
    \int_0^\tau dt\,
    \left[
        g^2(t)+G_{\mathrm{CD}}^2(t)
    \right].
    \label{eq:sta_dissipative_cost}
\end{eqnarray}
Figure~\ref{fig:q_heatmap} shows the dependence of the error $\chi_S$ on the
correlation time $\tau_c$ and the noise strength $\gamma$. The dashed line indicates the
threshold $\tau^2\gamma/\tau_c=10^{-2}$, with the region satisfying
$\tau^2\gamma/\tau_c<10^{-2}$ corresponding to the perturbative
regime. Note that Eq.~\eqref{eq:condition} still emerges as the condition for validity of our perturbative noise treatment, as can be seen by comparing the fluctuations to the maximum intensity of the {\it counterdiabatic} field $G_{\rm CD}$. We again use a fifth-order polynomial ramp for $g(t)$.

\begin{figure}[t]
    \centering
    \includegraphics[width=0.95\linewidth]{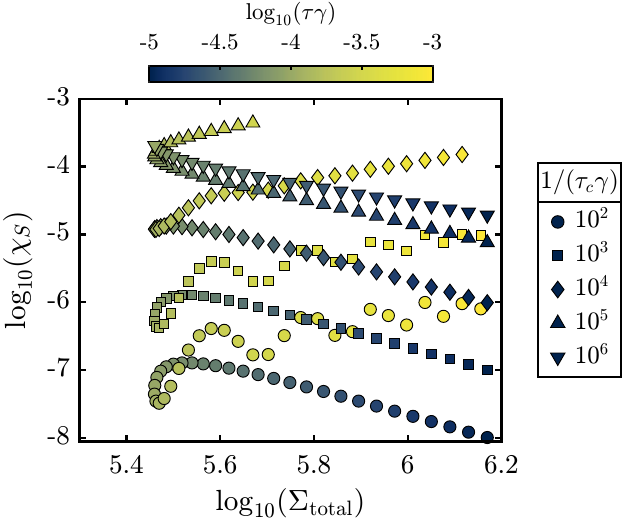}
    \caption{Scatter plot of infidelity $\chi_S$ vs energetic cost $\Sigma_{\rm total}$. Marker colour shows the operation time $\tau$, and different inverse noise  correlation times $1/\tau_c$ are represented by markers shown in legend. The data presented is cut-off for $\tau$ values not satisfying the perturbative boundary condition.}
    \label{fig:q_sigma_tradeoff}
\end{figure}

In Fig.~\ref{fig:q_sigma_tradeoff}, we show a log-scale scatter plot of $\chi_S$ versus the energy cost of the control $\Sigma_{\rm total}$. We can see two distinct regimes, depending on the duration of the protocol. For short protocol durations (blue points), we observe a trade-off behavior between the fidelity of the process and the energetic cost of the control. In other words, we have that the entropy production
 $\Sigma_{\rm total}$ increases as the infidelity $\chi_S$ decreases.

In contrast, for long protocol durations (yellow points), such a trade-off is not exhibited,

as both the infidelity and the energetic cost increase together.
This behavior is due to the fact that the entropy production $\Sigma_{\rm total}$ is dominated by the $\Sigma_z$ contribution in the long-duration regime. The cost of maintaining such a field increases with time, while the detrimental effect of noise on the system also becomes stronger. The oscillatory behavior in this regime varies with the inverse noise correlation time. Larger values of the inverse noise correlation time are excluded because they lie outside the perturbative regime. 

\begin{figure}
    \centering
    \includegraphics[width=0.95\linewidth]{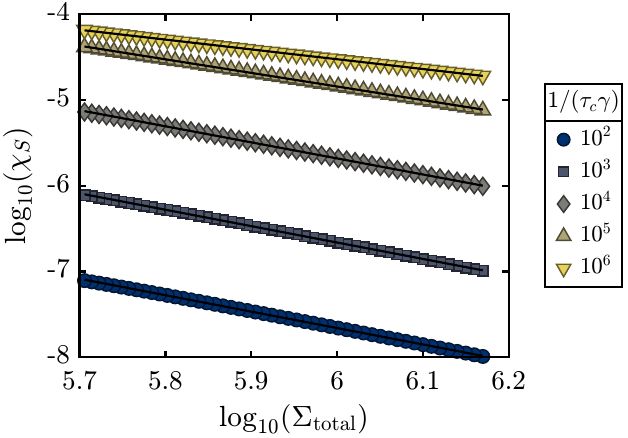}
    \caption{Short time branch of Fig. \ref{fig:q_sigma_tradeoff}. A power law fit (black lines) is outlined in Table \ref{tab:powerlaw_fit}.
    }
    \label{fig:powerlaw_fit_sigma_q}
\end{figure}

\begin{table}[t]
    \centering
    \caption{Power-law fit parameters for
    $\chi_S=A(\Sigma_{\mathrm{total}})^{\alpha}$
    at different inverse noise correlation times.}
    \label{tab:powerlaw_fit}

    \begin{ruledtabular}
    \begin{tabular}{cccc}
        $1/\tau_c$ & $\alpha$ & $A$ & $R^2$ \\
        \hline
        $10^{-2}$ & $-1.917$ & $6.95\times10^{3}$ & $0.9998$ \\
        $10^{-1}$ & $-1.913$ & $6.55\times10^{4}$ & $0.9998$ \\
        $1$       & $-1.875$ & $3.70\times10^{5}$ & $0.9998$ \\
        $10$      & $-1.586$ & $4.73\times10^{4}$ & $0.9996$ \\
        $10^{2}$  & $-1.146$ & $2.25\times10^{2}$ & $0.9999$
    \end{tabular}
    \end{ruledtabular}
\end{table}

The regime of short protocol duration is investigated further in Fig.~\ref{fig:powerlaw_fit_sigma_q}, where the relation between $\chi_S$ and $\Sigma_{\rm total}$ is fitted by a power law. The fitted power-law exponent converges as the inverse noise correlation time increases, see Table \ref{tab:powerlaw_fit}. As discussed in the limiting-case analysis, this regime requires $\gamma\tau\ll1$ and lies close to the boundary of perturbative validity. This therefore constitutes a clear trade-off, i.e., decreasing error requires increased dissipation in the classical controller, a finding consistent with the existing literature~\cite{stevens2025,fauquenot2025,Stevens2026}.

\begin{figure}
\centering
\includegraphics[width=0.95\linewidth]{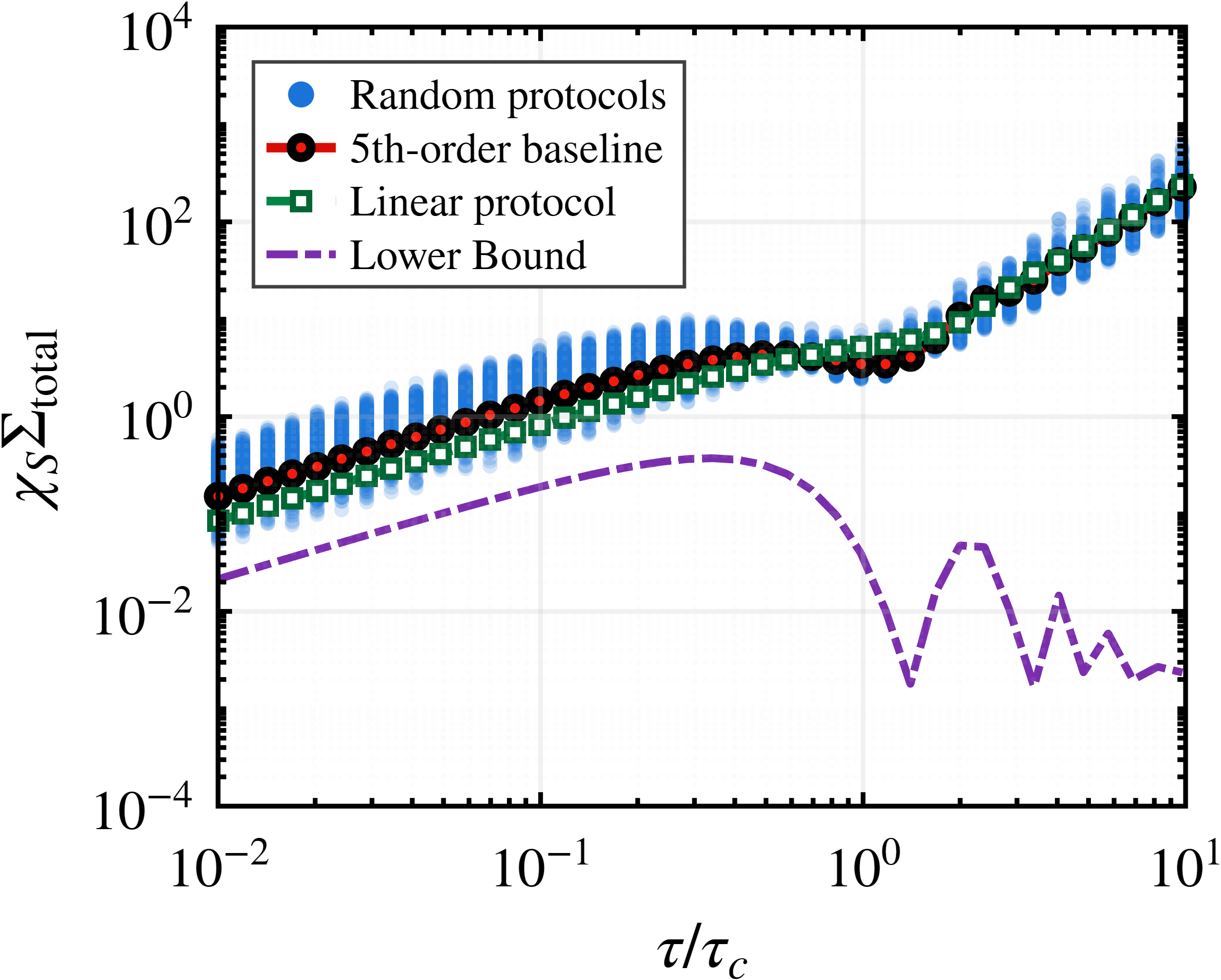}
\caption{Behavior of the product $ \chi_S \Sigma_{\mathrm{total}}$ as a function of the process time for an ensemble of randomly generated smooth monotonic protocols satisfying the same boundary conditions as the fifth-order polynomial baseline protocol. The red markers correspond to the baseline protocol, while the blue points represent the randomly generated protocols. The purple dashed line represents the lower bound given by the right-hand side of Eq.~\eqref{eq:qsigmamul} with fifth-order polynomial baseline. Although all sampled protocols lie above this curve, the bound is protocol-dependent. Note that the value of $\gamma$ determines the regime of validity.}
\label{fig:qSigma_tau_random_controls}
\end{figure}

In Fig.~\ref{fig:qSigma_tau_random_controls}, we investigate the behavior of the product $\chi_S\Sigma_{\mathrm{total}}$ in relation to the bound in Eq.~\eqref{eq:qsigmamul}. Both figures were obtained by generating an ensemble of smooth monotonic protocols based on the fifth-order polynomial baseline, all satisfying the same boundary conditions. Figure~\ref{fig:qSigma_tau_random_controls} corresponds to a scan over the protocol duration. These examples illustrate the limiting behavior of $\chi_S\Sigma_{\mathrm{total}}$ and show that, even under identical boundary conditions, different protocol shapes can
produce substantially different values of both $\chi_S$ and $\Sigma_{\mathrm{total}}$.

 Nevertheless, the product $\chi_S \Sigma_{\rm total}$ follows the same qualitative trend for all protocols. Figure~\ref{fig:qSigma_tau_random_controls} shows three distinct regimes: $\tau<\tau_c$, $\tau\approx\tau_c$, and $\tau>\tau_c$. In the first regime, $\chi_S$ decreases while $\Sigma_{\mathrm{total}}$ increases, leading to an approximately linear increase of the product $\chi_S\Sigma_{\mathrm{total}}$. Around $\tau\approx\tau_c$, the product becomes nearly constant. In the last regime, both $\chi_S$ and $\Sigma_{\mathrm{total}}$ increase simultaneously, resulting in a steeper slope. The main difference is that the present figure includes an ensemble of randomly generated smooth monotonic protocols. Although individual protocols exhibit some spread, they all follow the same qualitative behavior as the fifth-order baseline protocol.

\section{Conclusion and Outlook}
\label{sec:conclusions}
In this work, we have examined the link between entropy production and errors in quantum control, using a physical model of the intrinsically classical dissipative controller. Here, errors arise due to fluctuations of the classical variables generating the control fields, which are inevitably related to dissipation in the controller. Our approach thus captures a macroscopic contribution to the thermodynamic cost of quantum control, going beyond the mere energetics of the quantum system. 

Within this framework, we typically find a power-law tradeoff between error and entropy production, such that increasing fidelity comes at the cost of increased dissipation in the controller. In particular, we observe this behavior for gate fidelity in the commuting case (Fig.~\ref{fig:gate_infidelity_energy}) and also for the state fidelity of Landau-Zener protocols in the fast-driving (small $\tau$) limit (Fig.~\ref{fig:powerlaw_fit_sigma_q}) where the counterdiabatic contribution dominates. For slow LZ protocols, this tradeoff disappears as shown in Fig.~\ref{fig:q_sigma_tradeoff}. However, this slow-driving regime is less relevant because it leads to larger control errors, while fast control is generally desirable to improve resource efficiency and to avoid other sources of noise that our model does not explicitly account for. In general, minimizing the error-dissipation  product requires rapid control compared to the noise correlation time, as confirmed in Fig.~\ref{fig:qSigma_tau_random_controls} for a family of randomly generated STA protocols. 

There are several potential future directions building on this work. Here we consider a simple model of the classical controller. While more complex models could be considered, we speculate that our minimal model produces the least entropy and so gives a lower bound. One could also consider control fields which are even under time reversal such as electric fields, although we expect qualitatively similar results in this case. Our framework is also applicable beyond single-qubit examples, for example to control of ions in a Penning trap using a time-varying magnetic field \cite{kiely2015,masuda2015,Johnson2025}.

Another interesting line of inquiry concerns the role of fundamental speed or precision bounds on the classical controller. Previously, quantum speed limits have been related to the entropy production~\cite{deffner2017}. These approaches could potentially be extended to the case of a dissipative classical controller. Furthermore, the precision of currents in classical nonequilibrium systems is constrained by entropy production via thermodynamic uncertainty relations~\cite{Barato2015a, Gingrich2016, Horowitz2020}. Since fluctuations of the controller variables cause errors, this hints at a generic relationship between dissipation and control fidelity. Our framework provides a concrete setting in which these general questions could be investigated in future.

\begin{acknowledgments}
S.C.K. acknowledges financial support from TÜBİTAK-BİDEB through the 2214-A Yurt Dışı Doktora Sırası Araştırma Burs Programı (2214-A International Research Fellowship Programme for PhD Students). AK acknowledges financial support of Taighde \'Eireann – Research Ireland under grant number 24/PATH-S/12701. MTM is supported by a Royal Society University Research Fellowship. This project is co-funded by the European Union (Quantum Flagship project ASPECTS, Grant Agreement No.\ 101080167) and UK Research and Innovation (UKRI). Views and opinions expressed are however those of the authors only and do not necessarily reflect those of the European Union, Research Executive Agency or UKRI. Neither the European Union nor UKRI can be held responsible for them.

\end{acknowledgments}

\appendix

\section{Exact STA propagator and Heisenberg evolution}
\label{app:STA}

In this Appendix, we derive the exact shortcut-to-adiabaticity propagator and
the Heisenberg-picture connected correlation function used in the main text. The reference Landau--Zener Hamiltonian is
\begin{equation}
H_{\rm LZ}(t)
=
\Delta\sigma_x+g(t)\sigma_z.
\label{eq:LZ_bare_appendix}
\end{equation}
The counterdiabatic field $G_{CD}(t)$ is chosen to ensure that the state will mimic the adiabatic evolution of $H_{\rm LZ}$ for finite operation time $\tau$.

The instantaneous eigenvalues of $H_{\rm LZ}(t)$ are $E_{e,g}(t)=\pm E(t)$ with corresponding eigenvectors,
\begin{equation}
|n_e(t)\rangle
=
\begin{pmatrix}
\cos(\theta_t/2)\\[1mm]
\sin(\theta_t/2)
\end{pmatrix},
\qquad
|n_g(t)\rangle
=
\begin{pmatrix}
-\sin(\theta_t/2)\\[1mm]
\cos(\theta_t/2)
\end{pmatrix}.
\label{eq:LZ_eigenstates_appendix}
\end{equation}
Here we have defined the positive energy eigenvalue $E(t)=\sqrt{\Delta^2+g(t)^2}$ and the angle $\theta_t$ through
\begin{equation}
\sin\theta_t
=
x_t
=
\frac{\Delta}{E(t)},
\qquad
\cos\theta_t
=
z_t
=
\frac{g(t)}{E(t)}.
\label{eq:theta_components_appendix}
\end{equation}

For a general time-dependent Hamiltonian, the transitionless propagator may
be written as
\begin{eqnarray}
U(t)
=
e^{-i\alpha(t)}
|n_e(t)\rangle\langle n_e(0)|
+
e^{i\alpha(t)}
|n_g(t)\rangle\langle n_g(0)|, \nonumber \\
\label{eq:STA_spectral_propagator}
\end{eqnarray}

where we have assumed real eigenvectors such that $\langle n_{e,g}(t)|\dot n_{e,g}(t)\rangle=0$ and defined the dynamical phase $\alpha(t)=\int_0^t E(t')\,dt'$.

To express this in terms of Pauli rotations, we define a rotation about the $y$ axis $V(t)=e^{-i\theta_t\sigma_y/2}$, which gives $V(t)|0\rangle=|n_e(t)\rangle$ and $V(t)|1\rangle=|n_g(t)\rangle$. By also defining the diagonal matrix $D(t)=e^{-i\alpha(t)\sigma_z}$, the propagator can be written as
\begin{eqnarray}
U(t)
&=&
V(t)D(t)V^\dagger(0)  \\
&=& e^{-i\theta_t\sigma_y/2}
e^{-i\alpha(t)\sigma_z}
e^{i\theta_0\sigma_y/2}.
\label{eq:STA_rotation_decomposition}
\end{eqnarray}

\subsection{Heisenberg-picture operators as Bloch rotations}

For the independent-controller noise model, the perturbation to
the ideal Hamiltonian is
\begin{equation}
    \delta H(t)
    =
    \delta g_\lambda(t)W_\lambda
    +
    \delta g_\beta(t)W_\beta.
\end{equation}
The two perturbation operators can be written as
\begin{equation}
    W_a
    =
    \mathbf w_a\cdot\boldsymbol{\sigma},
    \qquad
    a\in\{\lambda,\beta\},
\end{equation}
with
\begin{equation}
    \mathbf w_\lambda
    =
    \begin{pmatrix}
        0\\
        0\\
        1
    \end{pmatrix},
    \qquad
    \mathbf w_\beta
    =
    \begin{pmatrix}
        0\\
        1\\
        0
    \end{pmatrix},
    \label{eq:independent_noise_vectors_appendix}
\end{equation}
and $\boldsymbol{\sigma}=(\sigma_x \, \, \sigma_y \, \, \sigma_z)^T$. Their Heisenberg-picture representations are
\begin{equation}
    W_a^H(t)
    =
    U^\dagger(t)W_aU(t).
\end{equation}

For a qubit unitary $U$ associated with the three-dimensional
rotation $R_U$, the Pauli-vector transformation satisfies
\begin{equation}
    U
    \left(
        \mathbf v\cdot\boldsymbol{\sigma}
    \right)
    U^\dagger
    =
    \left(
        R_U\mathbf v
    \right)\cdot\boldsymbol{\sigma},
\end{equation}
or, equivalently,
\begin{equation}
    U^\dagger
    \left(
        \mathbf v\cdot\boldsymbol{\sigma}
    \right)
    U
    =
    \left(
        R_U^{-1}\mathbf v
    \right)\cdot\boldsymbol{\sigma}.
    \label{eq:inverse_Bloch_rotation_appendix}
\end{equation}

Since $D(t)=e^{-i\alpha(t)\sigma_z}
=e^{-i[2\alpha(t)]\sigma_z/2}$, the accumulated
dynamical phase $\alpha(t)$ corresponds to a
Bloch-sphere rotation by an angle $2\alpha(t)$
about the $z$ axis. The rotation associated
with the propagator is therefore
\begin{equation}
    R_U(t)
    =
    R_y(\theta_t)
    R_z\!\left(2\alpha(t)\right)
    R_y(-\theta_0).
\end{equation}
It follows from
Eq.~\eqref{eq:inverse_Bloch_rotation_appendix} that the Bloch
vector of each Heisenberg-picture perturbation operator is
\begin{equation}
    \mathbf A_a(t)
    =
    R_y(\theta_0)
    R_z\!\left[-2\alpha(t)\right]
    R_y(-\theta_t)\mathbf w_a.
    \label{eq:independent_A_rotation_appendix}
\end{equation}
The rotation matrices are
\begin{equation}
    R_y(\theta)
    =
    \begin{pmatrix}
        \cos\theta&0&\sin\theta\\
        0&1&0\\
        -\sin\theta&0&\cos\theta
    \end{pmatrix},
\end{equation}
and
\begin{equation}
    R_z\!\left(2\alpha(t)\right)
    =
    \begin{pmatrix}
        \cos[2\alpha(t)]&-\sin[2\alpha(t)]&0\\
        \sin[2\alpha(t)]&\cos[2\alpha(t)]&0\\
        0&0&1
    \end{pmatrix}.
\end{equation}

We use the notation
\begin{equation}
    x_t=\sin\theta_t,
    \qquad
    z_t=\cos\theta_t,
\end{equation}
where
\begin{equation}
    x_t
    =
    \frac{\Delta}{\sqrt{\Delta^2+g^2(t)}},
    \qquad
    z_t
    =
    \frac{g(t)}{\sqrt{\Delta^2+g^2(t)}}.
\end{equation}
We additionally define
\begin{equation}
    c_t
    =
    \cos\left[2\alpha(t)\right],
    \qquad
    s_t
    =
    \sin\left[2\alpha(t)\right].
\end{equation}

For the $\lambda$ channel, the first rotation gives
\begin{equation}
    R_y(-\theta_t)\mathbf w_\lambda
    =
    \begin{pmatrix}
        -x_t\\
        0\\
        z_t
    \end{pmatrix}.
\end{equation}
The subsequent $z$ rotation gives
\begin{equation}
    R_z\!\left[-2\alpha(t)\right]
    R_y(-\theta_t)\mathbf w_\lambda
    =
    \begin{pmatrix}
        -x_tc_t\\
        x_ts_t\\
        z_t
    \end{pmatrix}.
\end{equation}
Finally, applying $R_y(\theta_0)$ yields
\begin{equation}
    W_\lambda^H(t)
    =
    A_{\lambda x}(t)\sigma_x
    +
    A_{\lambda y}(t)\sigma_y
    +
    A_{\lambda z}(t)\sigma_z,
\end{equation}
where
\begin{align}
    A_{\lambda x}(t)
    &=
    x_0z_t-z_0x_tc_t,
    \label{eq:A_lambda_x_appendix}
    \\
    A_{\lambda y}(t)
    &=
    x_ts_t,
    \label{eq:A_lambda_y_appendix}
    \\
    A_{\lambda z}(t)
    &=
    z_0z_t+x_0x_tc_t.
    \label{eq:A_lambda_z_appendix}
\end{align}

For the $\beta$ channel,
\begin{equation}
    R_y(-\theta_t)\mathbf w_\beta
    =
    \begin{pmatrix}
        0\\
        1\\
        0
    \end{pmatrix},
\end{equation}
and hence
\begin{equation}
    R_z\!\left[-2\alpha(t)\right]
    R_y(-\theta_t)\mathbf w_\beta
    =
    \begin{pmatrix}
        s_t\\
        c_t\\
        0
    \end{pmatrix}.
\end{equation}
The final rotation gives
\begin{equation}
    W_\beta^H(t)
    =
    A_{\beta x}(t)\sigma_x
    +
    A_{\beta y}(t)\sigma_y
    +
    A_{\beta z}(t)\sigma_z,
\end{equation}
where
\begin{align}
    A_{\beta x}(t)
    &=
    z_0s_t,
    \label{eq:A_beta_x_appendix}
    \\
    A_{\beta y}(t)
    &=
    c_t,
    \label{eq:A_beta_y_appendix}
    \\
    A_{\beta z}(t)
    &=
    -x_0s_t.
    \label{eq:A_beta_z_appendix}
\end{align}

\subsection{Independent-channel noise kernels}

We assume that the two controller fluctuations are statistically
independent. Their correlations therefore satisfy
\begin{align}
    \avg{\delta g_\lambda(t)\delta g_\lambda(s)}
    &=
    C_\lambda(t,s),
    \\
    \avg{\delta g_\beta(t)\delta g_\beta(s)}
    &=
    C_\beta(t,s),
    \\
    \avg{\delta g_\lambda(t)\delta g_\beta(s)}
    &=
    \avg{\delta g_\beta(t)\delta g_\lambda(s)}
    =
    0.
    \label{eq:independent_classical_correlations_appendix}
\end{align}
In the calculations presented in the main text, the two
controllers are taken to have the same noise parameters, so that
\begin{equation}
    C_\lambda(t,s)
    =
    C_\beta(t,s)
    =
    C(t,s).
\end{equation}

Let the initial state be written in terms of its Bloch vector as
\begin{equation}
    \rho_0
    =
    \frac{1}{2}
    \left(
        \mathbb I+\mathbf r\cdot\boldsymbol{\sigma}
    \right).
\end{equation}
For each channel, we write
\begin{equation}
    W_a^H(t)
    =
    \mathbf A_a(t)\cdot\boldsymbol{\sigma},
    \qquad
    a\in\{\lambda,\beta\}.
\end{equation}
Using the Pauli identity
\begin{equation}
    \left(
        \mathbf a\cdot\boldsymbol{\sigma}
    \right)
    \left(
        \mathbf b\cdot\boldsymbol{\sigma}
    \right)
    =
    \left(
        \mathbf a\cdot\mathbf b
    \right)\mathbb I
    +
    i
    \left(
        \mathbf a\times\mathbf b
    \right)\cdot\boldsymbol{\sigma},
\end{equation}
the one-time expectation value is
\begin{equation}
    \avg{W_a^H(t)}_0
    =
    \mathbf r\cdot\mathbf A_a(t),
\end{equation}
while the two-time correlation function is
\begin{equation}
    \avg{W_a^H(t)W_a^H(s)}_0
    =
    \mathbf A_a(t)\cdot\mathbf A_a(s)
    +
    i\mathbf r\cdot
    \left[
        \mathbf A_a(t)\times\mathbf A_a(s)
    \right].
\end{equation}
The connected quantum kernel associated with channel $a$ is
therefore
\begin{align}
    K_S^{(a)}(t,s)
    ={}&
    \mathbf A_a(t)\cdot\mathbf A_a(s)
    -
    \left[
        \mathbf r\cdot\mathbf A_a(t)
    \right]
    \left[
        \mathbf r\cdot\mathbf A_a(s)
    \right]
    \nonumber\\
    &+
    i\mathbf r\cdot
    \left[
        \mathbf A_a(t)\times\mathbf A_a(s)
    \right].
    \label{eq:channel_connected_correlator_appendix}
\end{align}
The total susceptibility is therefore
\begin{align}
    \chi_S
    ={}&
    \int_0^\tau dt
    \int_0^\tau ds\,
    C_\lambda(t,s)K_S^{(\lambda)}(t,s)
    \nonumber\\
    &+
    \int_0^\tau dt
    \int_0^\tau ds\,
    C_\beta(t,s)K_S^{(\beta)}(t,s).
    \label{eq:independent_susceptibility_appendix}
\end{align}

When both controllers have the same correlation function,
$C_\lambda(t,s)=C_\beta(t,s)=C(t,s)$, this reduces to
\begin{equation}
    \chi_S
    =
    \int_0^\tau dt
    \int_0^\tau ds\,
    C(t,s)
    \left[
        K_S^{(\lambda)}(t,s)
        +
        K_S^{(\beta)}(t,s)
    \right].
    \label{eq:independent_susceptibility_equal_noise_appendix}
\end{equation}
\subsubsection{Initial state as the ground state}

Consider the standard Landau--Zener sweep starting
from the ground state of $H_0(0)$, with
$g(0)=-g_{\max}$. Its initial state is
$\rho_0=[\mathbb{I}-x_0\sigma_x-z_0\sigma_z]/2$. Therefore, the Bloch vector is $\mathbf r_g=(-x_0,0,-z_0)$, with $x_0^2+z_0^2=1$.

Substituting this vector into Eq. (A42), and using
the components of $\mathbf A_\lambda$ and
$\mathbf A_\beta$ from Eqs. (A24)--(A32), gives
\begin{align}
    \mathbf r_g\cdot\mathbf A_\lambda(t)
    &=-z_t,
    &
    \mathbf A_\lambda(t)\cdot\mathbf A_\lambda(s)
    &=z_tz_s+x_tx_s(c_tc_s+s_ts_s),
    \\
    \mathbf r_g\cdot\mathbf A_\beta(t)
    &=0,
    &
    \mathbf A_\beta(t)\cdot\mathbf A_\beta(s)
    &=c_tc_s+s_ts_s.
\end{align}

The cross-product terms are
\begin{align}
    \mathbf r_g\cdot
    [\mathbf A_\lambda(t)\times\mathbf A_\lambda(s)]
    &=-x_tx_s(s_tc_s-c_ts_s),
    \\
    \mathbf r_g\cdot
    [\mathbf A_\beta(t)\times\mathbf A_\beta(s)]
    &=-(s_tc_s-c_ts_s).
\end{align}

Hence, the connected channel kernels are
\begin{align}
    K_S^{(\lambda)}(t,s)
    &=
    x_tx_s
    \left[
        c_tc_s+s_ts_s-i(s_tc_s-c_ts_s)
    \right],
    \\
    K_S^{(\beta)}(t,s)
    &=
    c_tc_s+s_ts_s-i(s_tc_s-c_ts_s).
\end{align}

The imaginary part is antisymmetric under the exchange
$t\leftrightarrow s$. Since the classical noise correlation
functions are real and symmetric, the imaginary part vanishes
after integration over $[0,\tau]\times[0,\tau]$. The relevant
real channel kernels are consequently
\begin{align}
    K_S^{(\lambda)}(t,s)
    &=x_tx_s(c_tc_s+s_ts_s),
    \\
    K_S^{(\beta)}(t,s)
    &=c_tc_s+s_ts_s.
\end{align}

For identical noise-correlation functions in the two channels,
the ground-state susceptibility follows as
\begin{equation}
    \chi_S^{(g)}
    =
    \int_0^\tau dt\int_0^\tau ds\,
    C(t,s)(1+x_tx_s)(c_tc_s+s_ts_s).
    \label{eq:ground_state_susceptibility_appendix}
\end{equation}

For different channel correlations, the two kernels
are weighted separately by their respective correlation
functions, as in Eq.~\eqref{eq:independent_susceptibility_appendix}.

\subsubsection{Initial state as $\ket{0}$}
In the main text, we present the results for the initial ground state. Here, we additionally consider the computational-basis state $\ket{0}$ to illustrate the applicability of our framework beyond ground-state dynamics. Since this state is a superposition of the initial energy eigenstates, a relative dynamical phase accumulates between them, making the ideal final state dependent on the protocol duration. Related approaches can be found in the quantum-control literature, particularly in quantum-gate implementations, where control parameters determine the resulting unitary transformations rather than necessarily implementing a fixed state-transfer task \cite{Cao2013,Danin2024}. 

For this initial state, $\rho_0=\ket{0}\bra{0}$, the corresponding Bloch vector is $\mathbf r=(0,0,1)$. Substituting this into the general expression for the connected correlation function, the channel kernel becomes
\begin{align}
    K_S^{(a)}(t,s)
    ={}&
    A_{ax}(t)A_{ax}(s)
    +
    A_{ay}(t)A_{ay}(s)
    \nonumber\\
    &+
    i
    \left[
        A_{ax}(t)A_{ay}(s)
        -
        A_{ay}(t)A_{ax}(s)
    \right].
    \label{eq:channel_zero_state_correlator_appendix}
\end{align}
As in the ground-state case, the imaginary contribution
vanishes upon integration against the real, symmetric
noise correlation functions. The relevant real
channel kernels are therefore
\begin{equation}
    K_S^{(a)}(t,s)
    =
    A_{ax}(t)A_{ax}(s)
    +
    A_{ay}(t)A_{ay}(s).
    \label{eq:real_channel_kernel_appendix}
\end{equation}

\subsubsection{Gate fidelity for LZ model}
For the gate-fidelity kernel, the Pauli algebra together with
$\Tr(\sigma_i)=0$ and $\Tr(\mathbb{I})=d$ gives, for each channel,
\begin{equation}
    K_U^{(a)}(t,s)
    =
    \mathbf A_a(t)\cdot\mathbf A_a(s),
    \qquad
    a\in\{\lambda,\beta\}.
\end{equation}
For identical noise-correlation functions in the two independent
channels, the total kernel is
\begin{equation}
    K_U(t,s)
    =
    \mathbf A_\lambda(t)\cdot\mathbf A_\lambda(s)
    +
    \mathbf A_\beta(t)\cdot\mathbf A_\beta(s).
    \label{eq:independent_gate_kernel_appendix}
\end{equation}
\subsubsection{Lower bound with independent channels}
For independent controller fluctuations, we apply
Eq.~\eqref{eq:qsigmamul} separately to the two control channels,
\begin{equation}
    \frac{\chi_{S,a}\Sigma_a}{k_{\mathrm B}}
    \geq
    \frac{\theta_a^2\tau}{\tau_c}
    (\Delta Z_a)_0^2,
    \qquad a\in\{\lambda,\beta\}.
\end{equation}
Summing the two inequalities gives
\begin{equation}
    \frac{
        \chi_{S,\lambda}\Sigma_\lambda
        +\chi_{S,\beta}\Sigma_\beta
    }{k_{\mathrm B}}
    \geq
    \frac{\tau}{\tau_c}
    \left[
        \theta_\lambda^2(\Delta Z_\lambda)_0^2
        +
        \theta_\beta^2(\Delta Z_\beta)_0^2
    \right],
    \label{eq:independent_channel_bound_appendix}
\end{equation}
where $\Sigma_\lambda=\Sigma_z$ and
$\Sigma_\beta=\Sigma_{\mathrm{CD}}$ in the main text. Since $\chi_S=\chi_{S,\lambda}+\chi_{S,\beta}$ and
$\Sigma_{\mathrm{total}}=\Sigma_\lambda+\Sigma_\beta$,
all contributions are nonnegative and therefore
\begin{equation}
    \chi_S\Sigma_{\mathrm{total}}
    \geq
        \chi_{S,\lambda}\Sigma_\lambda
        +\chi_{S,\beta}\Sigma_\beta
    .
\end{equation}
Thus, the right-hand side of
Eq.~\eqref{eq:independent_channel_bound_appendix}
also lower bounds $\chi_S\Sigma_{\mathrm{total}}/k_{\mathrm B}$,
the quantity plotted in Fig.~\ref{fig:qSigma_tau_random_controls}. Note that for $g(0)=-g_{\max}$ and $g(\tau)=+g_{\max}$, we have $\theta_\beta=-\arctan(g_{\rm max}/\Delta)$.

\bibliography{draft/export.bib,draft/aexport_old.bib}

\end{document}